\PassOptionsToPackage{square,numbers}{natbib} 
\documentclass[10pt]{article}

\usepackage[utf8]{inputenc}
\usepackage[T1]{fontenc}
\usepackage[english]{babel}
\usepackage{graphicx}
\usepackage{amsmath,amssymb,amsbsy,amsthm,color,natbib}
\usepackage{multicol}
\usepackage{color} 
\usepackage{natbib}
\usepackage[linktoc=all,colorlinks=true,citecolor=blue,linkcolor=blue]{hyperref}
\usepackage[paperwidth=18cm,paperheight=29.7cm,hmargin=2cm,vmargin=2.5cm]{geometry}
\usepackage{authblk}
\usepackage{comment}
\usepackage{appendix}
\usepackage{bm}
\usepackage{booktabs}
\usepackage{tabularx}

\graphicspath{{Figures/}}

\title{A simple chaos indicator based on the Lagrangian \\ descriptor difference of neighboring orbits}

\author[1]{Javier Jiménez-López\thanks{javiej09@ucm.es (Corresponding author)}}
\author[2]{V. J. García-Garrido\thanks{vjose.garcia@uah.es}}
\affil[1]{Departamento de F\'isica Te\'orica, Universidad Complutense de Madrid, E-28040 Madrid, Spain}
\affil[2]{Departamento de F\'isica y Matem\'aticas, Facultad de Ciencias, Universidad de Alcal\'a, 28805 Alcal\'a de Henares, Madrid, Spain.}

\begin{document}

\maketitle


\begin{abstract}

In this paper we introduce a chaos indicator derivable from Lagrangian descriptors (LDs), defined as the difference in LD values between two neighboring trajectories. This \emph{difference LD} is remarkably easy to implement and interpret, offering a direct and intuitive measure of dynamical behavior. We provide a heuristic argument linking its growth to the regularity or chaoticity due to the underlying initial condition, considering the arclength-based formulation of LDs. To evaluate its effectiveness, we benchmark it against more elaborate LD-based chaos indicators and the Smaller Aligment Index (SALI) using two prototypical systems: the H\'enon-Heiles system and the Chirikov Standard Map. Our results show that, despite its simplicity, the difference LD matches the accuracy of more sophisticated methods, making it a robust and accessible tool for chaos detection in dynamical systems.

\end{abstract}

\noindent \textbf{keywords:} Chaos indicator, Lagrangian descriptor, Hamiltonian dynamics, Dynamical system.


\section{Introduction} \label{sec:Introduction}

One of the central aims of dynamical systems theory is to understand how trajectories evolve over time, particularly in distinguishing predictable, structured behavior from chaos. This involves determining the circumstances under which a system exhibits regular motions---such as periodic or quasi-periodic trajectories---as opposed to chaotic dynamics, where small variations in initial conditions can cause significant divergences in long-term behavior. Gaining insight into these contrasting regimes is essential: regular dynamics allow for reliable forecasting, whereas chaotic systems are inherently limited in their predictability.

The chaotic behavior of dynamical systems is not only of theoretical interest but also manifests itself in a wide range of relevant physical systems. In celestial mechanics, navigation satellites undergo chaotic transport due to the effect of gravitational perturbations \cite{Daquin2016, Daquin2019}, while the orbits of stars in barred galaxy models can switch from regular confinement to chaotic scattering depending on the physical parameters of the galactic potential considered \cite{Zotos2017}. At the laboratory scale, electromechanical setups powered by different sine, square  or triangular wave signals exhibit chaotic regimes \cite{Amarot2025}, and adaptive lattice models in condensed matter physics manifest order-chaos transitions as the coupling parameters are varied \cite{Sinha1995}. Chaos is also present in high-energy and nuclear physics, where the analysis of the double magic nucleus $^{208}\text{Pb}$ unveiled spectral signatures of chaos intermixed with regular shell-model behavior \cite{Dietz2017}, and studies of critical points in first-order quantum phase transitions have demonstrated how chaotic structures emerge in quantum systems \cite{Macek2011}. Furthermore, semiconductor microcavities operating in their optical modes \cite{Eleuch2012}, nonlinear dynamics in circular particle accelerators \cite{montanari2025}, and even the light-matter interaction in cavity QED setups \cite{Bastarrachea2017} or the non-perturbative regimes of Quantum Field Theory \cite{sonnenschein2025} are now being evaluated through the lens of chaos theory.

In order to properly characterize the dynamical behavior of a trajectory in phase space, one needs to use the so-called chaos indicators. Those that rely on the computation of the variational equations, such as the maximal Lyapunov Exponent (mLE) \cite{Skokos2010a}, the Fast Lyapunov Indicator (FLI) \cite{Froeschle1997a,Froeschle1997b}, the Relative Lyapunov Indicator (RLI) \cite{Sandor2004}, the Smaller Alignment Index (SALI) \cite{Skokos2001,Skokos2003,Skokos2004}, or its generalization, the Generalized Alignment Index (GALI) \cite{Skokos2008,moges2025}, which has been shown to be computable using the Multi-Particle method   \cite{MANYMANDA2025}, where instead of evolving deviation vectors nearby orbits are evolved, surpassing the computation of the variational equations, provide a perfect identification of the trajectory's nature. However, these chaos indicators are computationally expensive and relying on the variational equations can be problematic, as for some systems they can be highly complex \cite{Hillebrand2019}. Aiming to tackle these issues, chaos indicators derived from Lagrangian descriptors \cite{Hille22,zimper23,Daquin2022} have been recently introduced, along with implementations using differential algebra \cite{caliman2025} to further improve the results. In this framework, it is not necessary to obtain the variational equations as it is a trajectory-based methodology \cite{mancho2013,lopesino2017}, meaning that the computation of the indicator is achieved by adding one additional differential equation to the system composed of Hamilton's equations of motion, so that the computation time and mathematical complexity are considerably reduced. In this paper, we have developed a chaos indicator that can be derived in the mathematical framework of Lagrangian descriptors, obtained as the difference in the LD of two neighboring trajectories. As it only requires the integration of one neighboring initial condition, computing time can be further lowered, allowing the generation of large, high-quality datasets that facilitate the investigation of high-dimensional systems.

This paper is organized as follows. Section~\ref{sec:Methodology} begins with a brief overview of the method of Lagrangian descriptors (LDs), along with a review of various chaos indicators introduced in the literature based on this framework. We then introduce the \emph{difference LD}, denoted by $\Delta \mathcal{L}$ and provide a heuristic justification for its effectiveness in distinguishing between regular and chaotic initial conditions based on its growth behavior. In Section~\ref{sec:Results}, we present the main findings of our study. We compare the performance of $\Delta \mathcal{L}$ with other LD-based indicators and validate our results using SALI, a widely accepted benchmark for chaos detection. This validation is carried out using two well-known dynamical systems: the Hénon–Heiles system and the Chirikov Standard Map. Finally, Section~\ref{sec:Conclusions} summarizes the conclusions of our work.


\section{Lagrangian descriptors} 
\label{sec:Methodology}

The method of Lagrangian descriptors (LDs) is a trajectory-based diagnostic technique from Dynamical Systems Theory \cite{Garcia2022a} that was originally developed in the field of Geophysics to analyze Lagrangian transport and mixing processes in the ocean and the atmosphere \cite{madrid2009,mendoza10}.

Given a continuous dynamical system:
\begin{equation}
    \dfrac{d\mathbf{x}}{dt} = \mathbf{f}(\mathbf{x},t) \;,
    \label{general_ds}
\end{equation}
where the state vector $\mathbf{x} \in \mathcal{S}$ belongs to an $n$-dimensional phase space $\mathcal{S}$, a Lagrangian descriptor is a scalar function constructed by defining a non-negative function $\mathcal{F}(\mathbf{x}(t;\mathbf{x}_0),t)$ that depends on the initial condition $\mathbf{x}_0$ at time $t=t_0$. To determine the LD scalar field, denoted by $\mathcal{L}$, we set an integration time $\tau > 0$ and calculate:
\begin{equation}
    \mathcal{L}(\mathbf{x}_0,t_0,\tau) = \mathcal{L}_f(\mathbf{x}_0,t_0,\tau) + \mathcal{L}_b(\mathbf{x}_0,t_0,\tau) \;,
    \label{ld_def}
\end{equation}
where the forward ($\mathcal{L}_f$) and backward ($\mathcal{L}_b$) components of the LD function are given by:
\begin{equation}
\mathcal{L}_f(\mathbf{x}_0,t_0,\tau) = \int_{t_0}^{t_0+\tau} \mathcal{F}(\mathbf{x}(t;\mathbf{x}_0),t) \, dt \quad,\quad
\mathcal{L}_b(\mathbf{x}_0,t_0,\tau) = \int_{t_0-\tau}^{t_0} \mathcal{F}(\mathbf{x}(t;\mathbf{x}_0),t) \, dt \;. 
\end{equation}
As Eq. \eqref{ld_def} shows, the calculation of LDs involves the accumulation of the values taken by the function $\mathcal{F}$ along the trajectory starting at $\mathbf{x}_0$, as it evolves forward and backward in time. In the literature it has been shown rigorously that the scalar field generated by this method has the capability of identifying the invariant sets (equilibria, stable and unstable manifolds, tori, periodic orbits, etc.) that characterize the dynamical behavior of trajectories in the phase space of the system \cite{mancho2013,lopesino2017}. 

In this work we will focus on characterizing chaotic and regular behavior by means of the definition of LDs based on the arclength of trajectories, that is:
\begin{equation} \label{eq:DL_2}
    \mathcal{F}(\mathbf{x}(t;\mathbf{x}_0),t) = \lVert \mathbf{f}(\mathbf{x}(t;\mathbf{x}_0),t) \rVert = \sqrt{\sum_{i=1}^n f^{\,2}_i} \;,
\end{equation}
where $f_i$ is the $i$-th component of the vector field that determines the dynamical system in Eq. \eqref{general_ds}.

Lagrangian descriptors can also be used to explore the phase space structure of $d$-dimensional maps of the form:
\begin{equation}
\mathbf{x}_{n+1} = \mathbf{f}(\mathbf{x}_{n}) \;,\quad n=0,1,\ldots	\;.
\label{discrete_DS}
\end{equation}
Given $N>0$, which is the number of iterations both forward (using $\mathbf{f}$) and backward in time (applying $\mathbf{f}^{-1}$), we can define the discrete version of LDs \cite{Lopesino2015} as follows:
\begin{equation}
\mathcal{L}\left(\mathbf{x}_0,N\right) = \sum_{n=-N}^{N-1} \sqrt{\sum_{j = 1}^{d} \left(x^j_{n+1}-x^j_{n}\right)^{2}} \, .
\label{DLD}
\end{equation}
Note that Eq. \eqref{DLD} can be split into two different terms:
\begin{equation}
\mathcal{L}\left(\mathbf{x}_0,N\right) = \mathcal{L}_{f}\left(\mathbf{x}_0,N\right) + \mathcal{L}_{b}\left(\mathbf{x}_0,N\right) \;, 
\end{equation}
where $\mathcal{L}_{f}$ and $\mathcal{L}_{b}$ quantify, respectively, the contributions to the LDs of the forward and backward iterations of the orbit starting at the initial condition $\mathbf{x}_0$. This yields:
\begin{equation}
\mathcal{L}_f = \sum_{n=0}^{N-1} \sqrt{\sum_{j = 1}^{d} \left(x^j_{n+1}-x^j_{n}\right)^{2}} \quad,\quad \mathcal{L}_{b} = \sum_{n=-N}^{-1} \sqrt{\sum_{j = 1}^{d} \left(x^j_{n+1}-x^j_{n}\right)^{2}} \;.
\end{equation}

\subsection{Chaos indicators based on Lagrangian descriptors}

Chaos indicators derived from Lagrangian descriptors capable of distinguishing between regular and chaotic trajectories were initially proposed in \cite{Hille22,zimper23,caliman2025,Daquin2022}. More recently, they have been employed in Machine Learning frameworks, specifically to train Support Vector Machine classifiers that effectively identify order and chaos in Hamiltonian systems \cite{jimenez2025}. These LD-based techniques have been rigorously compared with SALI, a well-established benchmark for chaos detection, and against the Fast Lyapunov Indicator \cite{Daquin2022,caliman2025}, showing good agreement with this classical tool. Across all evaluations, LD-based indicators have consistently achieved accuracy rates exceeding $90\%$ relative to SALI, showing good agreement with what has been already presented in the literature \cite{zimper23}. A notable advantage of this approach lies in the simplicity of its implementation: computing the LD involves augmenting the system’s equations of motion with an additional differential (for flows) or difference (for maps) equation. This circumvents the need for solving the variational equations required by other indicators such as SALI, which track the evolution of deviation vectors. As a result, LD-based chaos detection methods offer reduced computational complexity and significantly lower CPU time demands.

In order to construct the chaos indicators from LDs, we need to obtain the neighbors of the initial condition that we would like to analyze. These neighbors are given by:
\begin{equation}
\mathbf{y}_i^{\pm} = \mathbf{x}_0 \pm \sigma_i \, \mathbf{e}_i \;\;,\;\; i=1,\ldots,n \;,
\end{equation}
where $\mathbf{e}_i$ is the $i$-th canonical basis vector in $\mathbb{R}^n$, and $\sigma_i$ represents the distance between the central point $\mathbf{x}_0$ (the initial condition we would like to classify) and its neighbors on the grid. The value of $n$ represents the dimension of the space where the initial conditions are selected. In this paper, since we are working with two-dimensional phase space slices, then $n = 2$, and thus each initial condition on the grid has 4 neighbors. Moreover, for our simulations we have chosen a value of $\sigma_i = 10^{-8}$, which ensures that the LD-based chaos indicators introduced below in Eq. \eqref{eq:chaos_inds} have enough accuracy for correctly identifying chaos and regularity when compared with SALI. An extensive study of how the spacing between the initial condition and its neighbors on the grid influence the accurate classification of chaotic and regular trajectories can be found in \cite{Hille22,zimper23}. Using these points, we can construct the following chaos indicators based on LDs:
\begin{equation}
\begin{split}
    D^n(\mathbf{x}_0) &= \dfrac{1}{2n \mathcal{L}_{f}(\mathbf{x}_0)} \sum_{i=1}^{n} \big|\mathcal{L}_{f}(\mathbf{x}_0)-\mathcal{L}_{f}\left(\mathbf{y}_i^{+}\right)\big|+\big|\mathcal{L}_{f}(\mathbf{x}_0)-\mathcal{L}_{f}\left(\mathbf{y}_i^{-}\right)\big| \;, \\  
    R^n(\mathbf{x}_0) &= \Bigg|1-\dfrac{1}{2n \mathcal{L}_{f}(\mathbf{x}_0)} \sum_{i=1}^{n} \mathcal{L}_{f}\left(\mathbf{y}_i^{+}\right)+\mathcal{L}_{f}\left(\mathbf{y}_i^{-}\right) \Bigg| \;, \\[.1cm]
    C^n(\mathbf{x}_0) &= \dfrac{1}{2n} \sum_{i=1}^{n} \dfrac{\big|\mathcal{L}_{f}\left(\mathbf{y}_i^{+}\right)-\mathcal{L}_{f}\left(\mathbf{y}_i^{-}\right)\big|}{\sigma_i} \;, \\[.1cm]
    S^n(\mathbf{x}_0) &= \dfrac{1}{n} \sum_{i=1}^{n} \dfrac{\big|\mathcal{L}_{f}\left(\mathbf{y}_i^{+}\right)-2\mathcal{L}_{f}(\mathbf{x}_0)+\mathcal{L}_{f}\left(\mathbf{y}_i^{-}\right)\big|}{\sigma_i^2} 
\end{split}
\label{eq:chaos_inds}
\end{equation}
where $\mathcal{L}_{f}(\cdot)$ is the forward LD value calculated for an integration time $\tau$. Note that the chaotic or regular nature of a trajectory is equivalently characterized if we integrate forward or backward in time, so only one of the components of the LD function is required to determine the chaos indicators.

\subsection{Characterizing chaos and regularity with $\Delta \mathcal{L}$ using the arclength definition} \label{subsec:arclength_proof}

In this section, we give the proof for our argument regarding the asymptotic behavior of the time evolution for the difference between the Lagrangian descriptor of two neighboring orbits depending on their nature.

For a given conservative, time-independent Hamiltonian function $\mathcal{H}$ with $n/2$ degrees of freedom, whose associated vector field is $\boldsymbol{f}(\boldsymbol{x})$, we consider an initial condition $\boldsymbol{x}_0 \in \mathbb{R}^{n}$ and a perturbed initial condition $\boldsymbol{x}_0 + \boldsymbol{\beta}$, where $\boldsymbol{\beta} \in \mathbb{R}^{n}$ and $\|\boldsymbol{\beta}\| \ll 1$. If we define the function $\Delta \mathcal{L}$ as the difference between the Lagrangian descriptor values (computed using the Euclidean norm) of the trajectories initialized at $\boldsymbol{x}_0$ and $\boldsymbol{x}_0 + \boldsymbol{\beta}$, the behavior of $\Delta \mathcal{L}$ as the trajectories evolve in phase space is as follows:

\begin{itemize}
    \item If $\boldsymbol{x}_0$ corresponds to a regular trajectory, then $\Delta \mathcal{L}$ is upper bounded by a function that grows linearly with time,
    \item If $\boldsymbol{x}_0$ corresponds to a chaotic trajectory, then $\Delta \mathcal{L}$ is upper bounded by a function that grows exponentially with time.
\end{itemize}

To begin the justification of our claim, we first consider the difference between the Lagrangian descriptor values of both trajectories, which is given by:
\begin{equation} \label{eq:diff_DL}
    \begin{aligned}
        \Delta \mathcal{L}(\boldsymbol{x}_0) = |\mathcal{L}(\boldsymbol{x}_0) - \mathcal{L}(\boldsymbol{x}_0 + \boldsymbol{\beta})| &= \left| \int_{t_0}^{t_0 + \tau}  \mathcal{F}(\boldsymbol{x}) \, dt \, - \int_{t_0}^{t_0 + \tau} \mathcal{F}(\boldsymbol{x^{\prime}}) \, dt \right| \,  \\
        &= \left| \int_{t_0}^{t_0 + \tau} \left[ \mathcal{F}(\boldsymbol{x}) \, - \, \mathcal{F}(\boldsymbol{x^{\prime}}) \right] \, dt \right| \, ,
    \end{aligned}
\end{equation}
where $\boldsymbol{x}^{\prime}$ is the trajectory generated by $\boldsymbol{x}_0 + \boldsymbol{\beta}$ and the function $\mathcal{F}(\boldsymbol{x})$ follows the definition given in Eq. \eqref{eq:DL_2}. As the perturbation $\boldsymbol{\beta}$ satisfies that $\|\boldsymbol{\beta}\| \ll 1$, we can expand $f_i(\boldsymbol{x}^{\prime})$ as a Taylor series up to first order:
\begin{equation}
    f_i(\boldsymbol{x^{\prime}}) \approx f_i(\boldsymbol{x}) + \nabla_x f_i(\boldsymbol{x}) \delta x_i(t) \, .
\end{equation}
with $\delta x_i(t)$ being the separation between both trajectories at time $t$. By squaring both sides of the last equation and neglecting second order terms, we get:
\begin{equation}
    f^{2}_i(\boldsymbol{x^{\prime}}) \approx f^{2}_i(\boldsymbol{x}) + 2 f_i(\boldsymbol{x}) \nabla_x f_i(\boldsymbol{x}) \,  \delta x_i (t) \, ,
\end{equation}

which can be reorganized to be:
\begin{equation}
    \sum_{i=1}^{n} f^{2}_i(\boldsymbol{x^{\prime}}) \approx \sum_{i=1}^{n} f^{2}_i(\boldsymbol{x}) + \varepsilon(\boldsymbol{x}, t) \, \, \, \, \text{with} \, \, \, \,  \varepsilon (\boldsymbol{x}, t) = 2 \boldsymbol{V}(\boldsymbol{x}) \cdot \delta \boldsymbol{x}(t) \, .
\end{equation}

where the dot product used to define $\varepsilon (\boldsymbol{x}, t)$ is $\boldsymbol{V}(\boldsymbol{x}) \cdot \delta \boldsymbol{x}(t) = \sum_{i=1}^{n} f_i(\boldsymbol{x}) \nabla_x f_i(\boldsymbol{x}) |_{x} \,  \delta x_i (t)$.

Defining the function $\mathcal{A}(\boldsymbol{x})$ as:
\begin{equation}
   \mathcal{A}(\boldsymbol{x}) = \sum_{i=1}^{n} f^{2}_i(\boldsymbol{x}) \, ,
\end{equation}
the terms inside the integral in Eq. \eqref{eq:diff_DL} can be written as:
\begin{equation}
    \sqrt{\mathcal{A}(\boldsymbol{x}) + \varepsilon(\boldsymbol{x}, t)} - \sqrt{\mathcal{A}(\boldsymbol{x})} \approx \dfrac{\varepsilon(\boldsymbol{x}, t)}{2\sqrt{\mathcal{A}(\boldsymbol{x})}} \, ,
\end{equation}
which substituted in Eq. \eqref{eq:diff_DL} gives:
\begin{equation}
    \begin{aligned}
        \Delta \mathcal{L} &\approx \left| \int_{t_0}^{t_0 + \tau} \dfrac{\varepsilon(\boldsymbol{x}, t)}{2\sqrt{\mathcal{A}(\boldsymbol{x})}} \, dt \right|= \left| \int_{t_0}^{t_0 + \tau} \dfrac{\varepsilon(\boldsymbol{x}, t)}{\sqrt{\mathcal{A}(\boldsymbol{x})}} \, dt \right| = \left| \int_{t_0}^{t_0 + \tau} \dfrac{\boldsymbol{V}(\boldsymbol{x}) \cdot \delta \boldsymbol{x}(t)}{\left( \sum_{i=1}^{n} f^{2}_i(\boldsymbol{x}) \right)^{1/2}} \, dt \right|  \\
        &  \, \leq \mathcal{C}_{\text{max}} \int_{t_0}^{t_0 + \tau} \| \delta \boldsymbol{x}(t) \| \, dt \, ,
    \end{aligned}
\end{equation}
where the constant $\mathcal{C}_{\text{max}}$ is defined as:

\begin{equation}
\mathcal{C}_{\text{max}} = \max_{t \in [t_0, t_0 + \tau]} \left[ \frac{\lVert \boldsymbol{V}(\boldsymbol{x}) \rVert}{\lVert \boldsymbol{f}(\boldsymbol{x}) \rVert} \right].
\end{equation}

The time-evolution of the separation between the trajectories is known to be given by \cite{strogatz2018nonlinear}:
\begin{equation}
    \|\delta \boldsymbol{x}(t) \| \approx \|\boldsymbol{\beta}\|\exp{(\lambda_{\text{max}} (t - t_0))} \, ,
\end{equation}
so that $\Delta \mathcal{L}$ can then be upper bounded as a function of $\|\boldsymbol{\beta}\|$, $\lambda_{\text{max}}$ and $\tau$:
\begin{equation}
    \Delta \mathcal{L} \leq  \mathcal{C}_{\text{max}} \|\boldsymbol{\beta}\| \int_{0}^{\tau} \exp{(\lambda_{\text{max}} u)} \, du \, .
\end{equation}

Considering that for a regular initial condition $\lambda_{\text{max}} =0$, the upper bound of $\Delta \mathcal{L}$ will be a function that grows linearly with the time of integration:
\begin{equation} \label{eq:DL_reg_bound}
    \Delta \mathcal{L}  \leq \mathcal{C}_{\text{max}} \|\boldsymbol{\beta}\| \tau \, ,
\end{equation}
and in the case of a chaotic initial condition, $\lambda_{\text{max}} > 0$, meaning that the behavior of $\Delta \mathcal{L}$ will be upper bounded by:
\begin{equation}
    \Delta \mathcal{L} \leq \mathcal{C}_{\text{max}} \|\boldsymbol{\beta}\| \dfrac{\exp{(\lambda_{\text{max}} \tau)} - 1}{\lambda_{\text{max}}} \, .
\end{equation}

In order to study the asymptotic behaviors of the bounds we have derived, it is convenient to introduce the auxiliary function $\Omega$, defined as the time average of the Lagrangian descriptor for the trajectory whose dynamical behavior we are interested in unveiling. Then:

\begin{equation}
    \Omega = \dfrac{\mathcal{L}_f}{\tau} \, ,
\end{equation}
which, for a regular orbit, is known to be asymptotically a constant \cite{doi:10.1142/S0218127417300014, NAIK2019104907, MONTES2021105860}, as it follows from  Birkhoff's Ergodic Partition Theorem \cite{10.1063/1.166399}.

Then, the upper bound that we have derived in Eq. (\ref{eq:DL_reg_bound}) is consistent with Birkhoff's Ergodic Partition Theorem for a regular orbit so that the asymptotic behavior of its average over time will be:

\begin{equation} \label{eq:Delta_L_avg}
    \dfrac{\Delta \mathcal{L}}{\tau} \sim \mathcal{C}_{\text{max}} \|\boldsymbol{\beta}\| \, .
\end{equation}

In the case of a chaotic trajectory, the asymptotic time evolution of $\Delta \mathcal{L}$ will follow the exponential law that is characteristic of chaotic dynamics:

\begin{equation}
     \Delta \mathcal{L} \sim \mathcal{C}_{\text{max}} \|\boldsymbol{\beta}\| \dfrac{\exp{(\lambda_{\text{max}} \tau)} - 1}{\lambda_{\text{max}}} \, .
\end{equation}

These asymptotic behaviors are depicted in Fig. \ref{fig:comparison_DL} for a regular and a chaotic orbit in the well-known H\'enon-Heiles system.


\section{Results} 
\label{sec:Results}

In order to illustrate the performance of the $\Delta \mathcal{L}$ indicator, we have chosen the H\'enon-Heiles Hamiltonian \cite{henon1964applicability}:
\begin{equation}
    \mathcal{H}(x,y,p_x,p_y) = \dfrac{p_x^2}{2} + \dfrac{p_y^2}{2} + \dfrac{1}{2}(x^2+y^2) + x^2y - \dfrac{1}{3}y^3 \, ,
    \label{eq:Ham_HH}
\end{equation}
with the Poincar\'e section defined as $x = 0$ and $p_x > 0$,  and the Chirikov Standard Map \cite{Chirikov1979}:
\begin{equation}
    \begin{cases}
        x_{n+1} = x_n + y_{n+1} \\[.1cm]
        y_{n+1} = y_n + \dfrac{K}{2\pi} \sin(2\pi x_n)
    \end{cases} (\text{mod } 1) \;,
    \label{eq.std_map}
\end{equation}
as they are well-know examples commonly used as benchmarking systems for chaos indicators. 

The generation of initial conditions for the H\'enon-Heiles system has been done by creating randomly generated sets of $10^{5}$ initial conditions for each energy value analyzed. Our choice of using a random sampling of the constant energy surface to generate initial conditions is motivated by the difficulty of extending the usual regular-grid approach to higher dimensional problems. In a $d$-dimensional space, a cartesian grid with $n$ points per coordinate contains $n^d$ nodes, which even for modest resolutions can become an astronomically large number of initial conditions that must be stored. This exponential growth has been already reported \cite{Bungartz_Griebel_2004} and is known as the \textit{curse of dimensionality} problem. On the other hand, a random sampling allows us to choose an arbitrary total number of initial conditions, independently of $d$, while exploring the constant-energy surface in a statistically homogeneous way. Then, the neighbor initial conditions were generated by setting a random direction in phase space given by the vector $\boldsymbol{\beta}$ and adding the condition $\|\boldsymbol{\beta}\| = \sigma = 10^{-8}$. The integration time selected for the calculation of the chaos indicators in this system was $10^{4}$ units of time as we have found this value to give consistent results while keeping a reasonable computation cost for the ODE solver given in \cite{prince1981}, which has provided a conservation of energy in all the integrations performed of, at least, $|\Delta \mathcal{H}| = |\mathcal{H}(t) - \mathcal{H}(t_0)| \leq 10^{-8}$. To carry out the integration of this datasets, we have used the Python package \textbf{Chaoticus} for parallel integration of continuous Hamiltonian systems \cite{jimenezlopez2025b}. In the case of the Chirikov Standard Map, we used a $320 \times 320$ regular grid of initial conditions for each value of $\mathrm{K}$ and the generation of the neighboring initial conditions was done using the same methodology as for the H\'enon-Heiles system, setting $\|\boldsymbol{\beta}\| = \sigma = 10^{-8}$. For this system, the number of steps simulated to calculate the chaos indicators given in Eq.(\ref{eq:chaos_inds}), Eq.(\ref{eq:diff_DL}) and for SALI was $10^{5}$.

The metrics that we have employed in order to compare the performance of $\Delta \mathcal{L}$ against other chaos indicators derived from Lagrangian descriptors, given in Eq.(\ref{eq:chaos_inds}), are the F1-Score \cite{vanrijsbergen1979,powers2020}:
\begin{equation} 
    \text{F1-Score} = \dfrac{2 \, \text{TP}}{2 \, \text{TP} + \text{FP} + \text{FN}} \;,
    \label{eq:metric}
\end{equation}
and the accuracy:
\begin{equation} 
    \text{Accuracy} = \dfrac{\text{TP} + \text{TN}}{\text{TP} + \text{TN} + \text{FP} + \text{FN}} \;,
    \label{eq:metric2}
\end{equation}
where TP stands for True Positives (chaotic trajectories that are classified as chaotic by $\Delta \mathcal{L}$), FP represents False Positives (regular trajectories that where classified as chaotic with $\Delta \mathcal{L}$) and FN stands for False Negatives (chaotic trajectories that where classified as regular by $\Delta \mathcal{L}$). The true nature of the trajectories was obtained using SALI \cite{Skokos2004}, which is considered to be a ground truth chaos indicator. The reason behind using the F1-Score combined with the accuracy to validate our results is that it provides a robust and balanced assessment, harmonizing the accuracy of chaotic predictions (precision) and the ability to identify the actual chaotic trajectories (recall), which ensures that $\Delta \mathcal{L}$ is not only conservative in its chaotic labeling but also consistent in the detection of true chaos. Furthermore, the F1-Score is particularly well-suited for working with imbalanced sets, which is highly relevant as the chaotic fraction of phase space might differ from the regular one. This makes the F1-Score more reliable and comprehensive than other metrics such as accuracy or precision, since it reflects the real effectiveness of $\Delta \mathcal{L}$ as a chaos indicator. However, if the number of chaotic initial conditions is very low, the F1-Score will have a value very close to zero even though the classification is being correctly done. To assess this issue, the accuracy is also employed as a metric. By combining the results given by both metrics, one can easily analyze the performance of $\Delta \mathcal{L}$.

The results obtained for the F1-Score in the H\'enon-Heiles system are given for $\Delta \mathcal{L}$ and the other chaos indicators derived from Lagrangian descriptors in Table \ref{tab:HH_F1} and for the accuracy in Table \ref{tab:HH_AC}; and for the Chirikov Standard Map in Table \ref{tab:SM_F1} and Table \ref{tab:SM_AC}, respectively. They show that the $\Delta \mathcal{L}$ indicator can perform as well as the other chaos indicators derived from Lagrangian descriptors, or even better, in almost all the test cases we have considered for our analysis.

\begin{table}[htbp]
\centering
\caption{Values for the F1-score metric defined in Eq.~\eqref{eq:metric} for the different chaos indicators based on Lagrangian descriptors, benchmarked against SALI, for the example cases shown in Fig.~\ref{fig:HH_results} for the H\'enon--Heiles system. The computations were done with $\|\boldsymbol{\beta}\| = \sigma = 10^{-8}$ for all cases. The values of the thresholds used can be found in Table~\ref{tab:HH_thres}.}
\label{tab:HH_F1}
\vspace{0.5cm}
\begin{tabular}{lcccc}
\toprule
\textbf{Indicator} &
$\boldsymbol{\mathcal{H}=1/8}$ &
$\boldsymbol{\mathcal{H}=1/10}$ &
$\boldsymbol{\mathcal{H}=1/12}$ &
$\boldsymbol{\mathcal{H}=1/15}$ \\
\midrule
$\mathcal{D}$        & 0.996703 & 0.606131 & 0.003717 & 0.000000 \\
$\mathcal{R}$        & 0.994953 & 0.904319 & 0.777401 & 0.285714 \\
$\mathcal{C}$        & 0.992020 & 0.935157 & 0.877093 & 0.104348 \\
$\mathcal{S}$        & 0.800434 & 0.310543 & 0.042935 & 0.000343 \\
$\Delta \mathcal{L}$ & 0.995976 & 0.962039 & 0.950513 & 0.833333 \\
\bottomrule
\end{tabular}
\end{table}

\begin{table}[htbp]
\centering
\caption{Values for the accuracy metric defined in Eq.~\eqref{eq:metric2} for the different chaos indicators based on Lagrangian descriptors, benchmarked against SALI, for the example cases shown in Fig.~\ref{fig:HH_results} for the H\'enon--Heiles system. The computations were done with $\|\boldsymbol{\beta}\| = \sigma = 10^{-8}$ for all cases. The values of the thresholds used can be found in Table~\ref{tab:HH_thres}.}
\label{tab:HH_AC}
\vspace{0.5cm}
\begin{tabular}{lcccc}
\toprule
\textbf{Indicator} &
$\boldsymbol{\mathcal{H}=1/8}$ &
$\boldsymbol{\mathcal{H}=1/10}$ &
$\boldsymbol{\mathcal{H}=1/12}$ &
$\boldsymbol{\mathcal{H}=1/15}$ \\
\midrule
$\mathcal{D}$        & 0.99491 & 0.99143 & 0.99905 & 0.99999 \\
$\mathcal{R}$        & 0.99360 & 0.98731 & 0.99852 & 0.99994 \\
$\mathcal{C}$        & 0.99593 & 0.98991 & 0.99920 & 1.00000 \\
$\mathcal{S}$        & 0.99365 & 0.98818 & 0.99851 & 0.99996 \\
$\Delta \mathcal{L}$ & 0.99532 & 0.98060 & 0.99889 & 0.99998 \\
\bottomrule
\end{tabular}
\end{table}

\begin{table}[htbp]
\centering
\caption{Values for the F1-score metric defined in Eq.~\eqref{eq:metric} for the different chaos indicators based on Lagrangian descriptors, benchmarked against SALI, for the example cases shown in Fig.~\ref{fig:SM_results} of the Chirikov Standard Map. The computations were done with $\|\boldsymbol{\beta}\| = \sigma = 10^{-8}$ for all cases. The values of the thresholds used can be found in Table~\ref{tab:SM_thres}.}
\label{tab:SM_F1}
\vspace{0.5cm}
\begin{tabular}{lccc}
\toprule
\textbf{Indicator} &
$\boldsymbol{\mathrm{K}=0.5}$ &
$\boldsymbol{\mathrm{K}=0.971635}$ &
$\boldsymbol{\mathrm{K}=1.5}$ \\
\midrule
$\mathcal{D}$        & 0.937335 & 0.989026 & 0.994423 \\
$\mathcal{R}$        & 0.922521 & 0.991406 & 0.994499 \\
$\mathcal{C}$        & 0.930732 & 0.989180 & 0.994334 \\
$\mathcal{S}$        & 0.934035 & 0.991932 & 0.996975 \\
$\Delta \mathcal{L}$ & 0.955283 & 0.993100 & 0.995263 \\
\bottomrule
\end{tabular}
\end{table}

\begin{table}[htbp]
\centering
\caption{Values for the accuracy metric defined in Eq.~\eqref{eq:metric2} for the different chaos indicators based on Lagrangian descriptors, benchmarked against SALI, for the example cases shown in Fig.~\ref{fig:SM_results} of the Chirikov Standard Map. The computations were done with $\|\boldsymbol{\beta}\| = \sigma = 10^{-8}$ for all cases. The values of the thresholds used can be found in Table~\ref{tab:SM_thres}.}
\label{tab:SM_AC}
\vspace{0.5cm}
\begin{tabular}{lccc}
\toprule
\textbf{Indicator} &
$\boldsymbol{\mathrm{K}=0.5}$ &
$\boldsymbol{\mathrm{K}=0.971635}$ &
$\boldsymbol{\mathrm{K}=1.5}$ \\
\midrule
$\mathcal{D}$        & 0.998145 & 0.989805 & 0.992012 \\
$\mathcal{R}$        & 0.997695 & 0.991982 & 0.992119 \\
$\mathcal{C}$        & 0.997920 & 0.989941 & 0.991885 \\
$\mathcal{S}$        & 0.997920 & 0.992471 & 0.995654 \\
$\Delta \mathcal{L}$ & 0.998682 & 0.993584 & 0.993213 \\
\bottomrule
\end{tabular}
\end{table}

\begin{table}[htbp]
\centering
\caption{Values for the thresholds used for the different chaos indicators based on Lagrangian descriptors for the example cases shown in Fig.~\ref{fig:HH_results} for the H\'enon--Heiles system. The computations were done with $\|\boldsymbol{\beta}\| = \sigma = 10^{-8}$ and the threshold for SALI was chosen to be $\log_{10}(\mathrm{SALI}_{\text{thres}}) = -8$ for the four cases.}
\label{tab:HH_thres}
\vspace{0.5cm}
\begin{tabular}{lcccc}
\toprule
\textbf{Indicator} &
$\boldsymbol{\mathcal{H}=1/8}$ &
$\boldsymbol{\mathcal{H}=1/10}$ &
$\boldsymbol{\mathcal{H}=1/12}$ &
$\boldsymbol{\mathcal{H}=1/15}$ \\
\midrule
$\log_{10}(\mathcal{D}_{\text{thres}})$              & $-6.0$ & $-6.0$ & $-6.0$ & $-6.0$ \\
$\log_{10}(\mathcal{R}_{\text{thres}})$              & $-7.5$ & $-8.0$ & $-7.5$ & $-7.5$ \\
$\log_{10}(\mathcal{C}_{\text{thres}})$              & $ 6.0$ & $ 5.0$ & $ 6.0$ & $ 6.0$ \\
$\log_{10}(\mathcal{S}_{\text{thres}})$              & $12.5$ & $12.5$ & $12.5$ & $12.5$ \\
$\log_{10}(\Delta \mathcal{L}_{\text{thres}})$       & $-2.5$ & $-3.0$ & $-3.0$ & $-2.0$ \\
\bottomrule
\end{tabular}
\end{table}

\begin{table}[htbp]
\centering
\caption{Values for the thresholds used for the different chaos indicators based on Lagrangian descriptors for the example cases shown in Fig.~\ref{fig:SM_results} of the Chirikov Standard Map. The computations were done with $\|\boldsymbol{\beta}\| = \sigma = 10^{-8}$ and the threshold for SALI was chosen to be $\log_{10}(\mathrm{SALI}_{\text{thres}}) = -13$ for the three cases.}
\label{tab:SM_thres}
\vspace{0.5cm}
\begin{tabular}{lccc}
\toprule
\textbf{Indicator} &
$\boldsymbol{\mathrm{K}=0.5}$ &
$\boldsymbol{\mathrm{K}=0.971635}$ &
$\boldsymbol{\mathrm{K}=1.5}$ \\
\midrule
$\log_{10}(\mathcal{D}_{\text{thres}})$        & $-2.5$ & $-3.25$ & $-2.0$ \\
$\log_{10}(\mathcal{R}_{\text{thres}})$        & $-3.0$ & $-4.5$  & $-3.0$ \\
$\log_{10}(\mathcal{C}_{\text{thres}})$        & $ 6.5$ & $ 6.0$  & $ 7.0$ \\
$\log_{10}(\mathcal{S}_{\text{thres}})$        & $11.0$ & $10.5$  & $12.0$ \\
$\log_{10}(\Delta \mathcal{L}_{\text{thres}})$ & $ 1.0$ & $ 0.0$  & $ 1.5$ \\
\bottomrule
\end{tabular}
\end{table}

\begin{figure}[htbp]
    \centering
    A) \includegraphics[scale = 0.42]{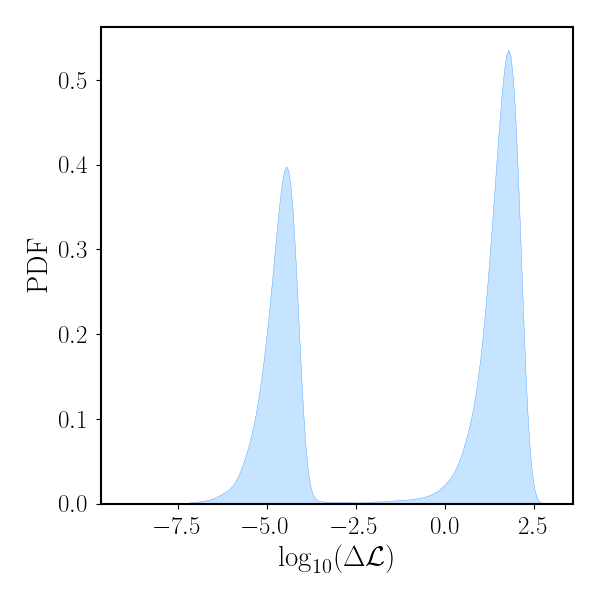}
    B) \includegraphics[scale = 0.42]{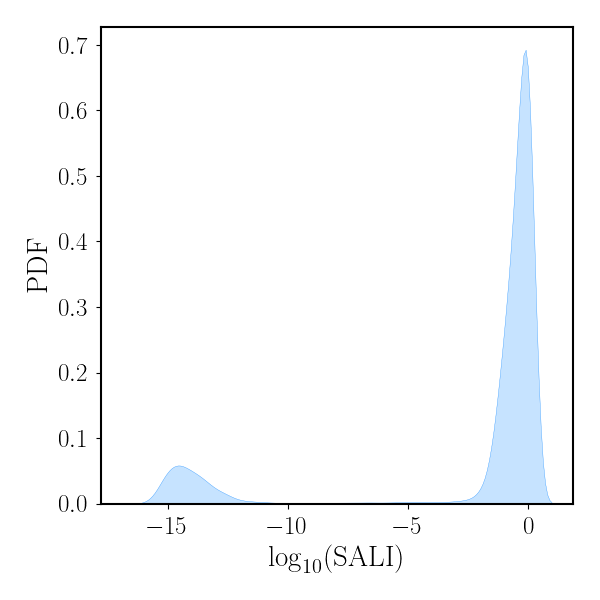} \\
    C) \includegraphics[scale = 0.42]{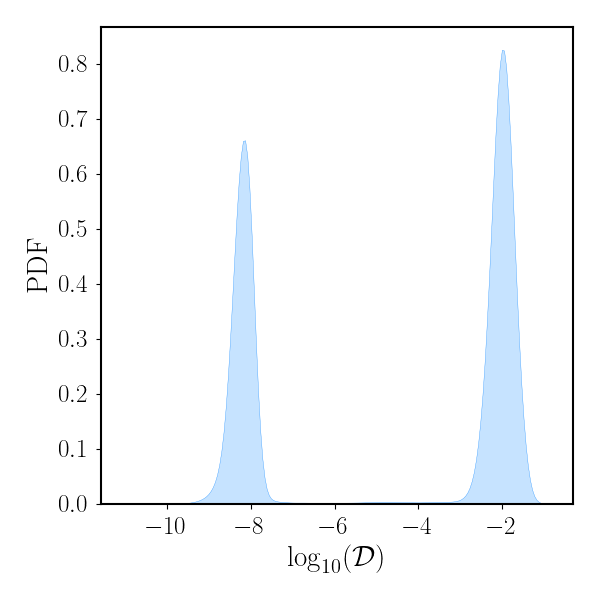}
    D) \includegraphics[scale = 0.42]{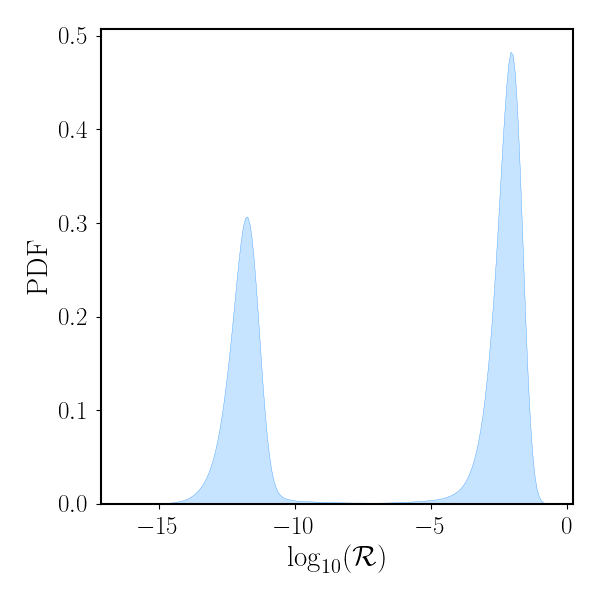} \\
    E) \includegraphics[scale = 0.42]{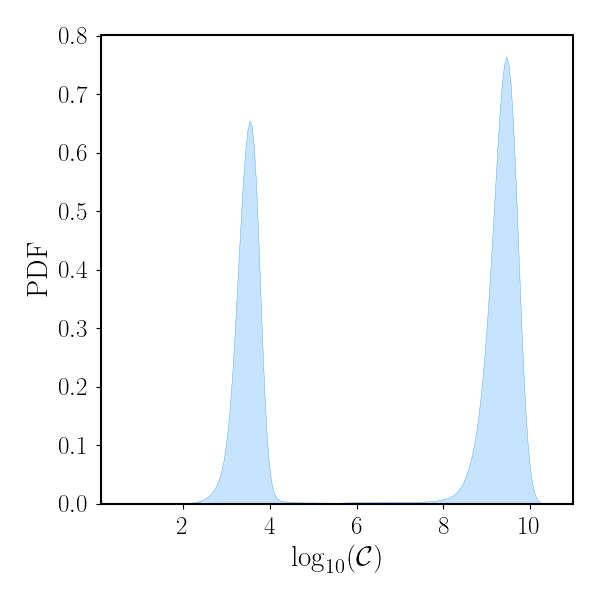}
    F) \includegraphics[scale = 0.42]{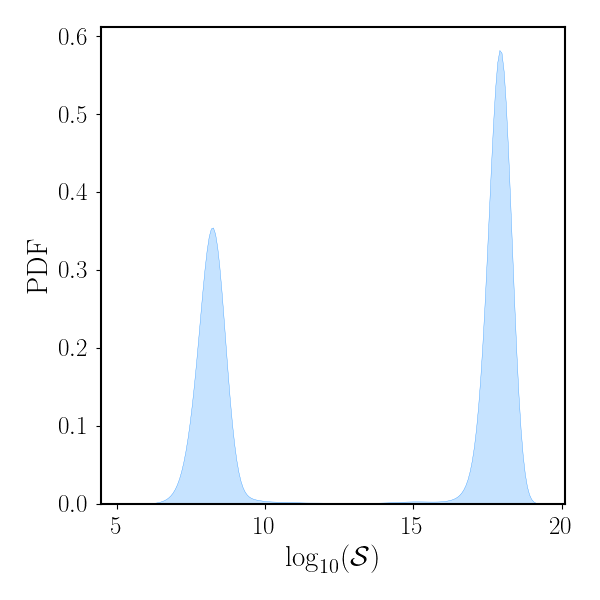}
    \caption{Histograms for the different chaos indicators considered in this analysis for the H\'enon-Heiles hamiltonian calculated with $10^{5}$ randomly generated initial conditions and $\mathcal{H} = 1/8$ integrated for $\tau = 10^{4}$ units of time and $\| \boldsymbol{\beta} \| = \sigma = 10^{-8}$. A) distribution of the $\log_{10}(\Delta \mathcal{L})$ indicator; B) distribution of the $\log_{10}(\text{SALI})$ indicator; C) distribution of the $\log_{10}(\mathcal{D})$; D) distribution of the $\log_{10}(\mathcal{R})$; E) distribution of the $\log_{10}(\mathcal{C})$; F) distribution of the $\log_{10}(\mathcal{S})$. Note that the indicators given in Eq. \eqref{eq:chaos_inds} were calculated with $n = 2$.}
    \label{fig:histograms}
\end{figure}

\begin{figure}[htbp]
    \centering
    A) \includegraphics[scale = 0.42]{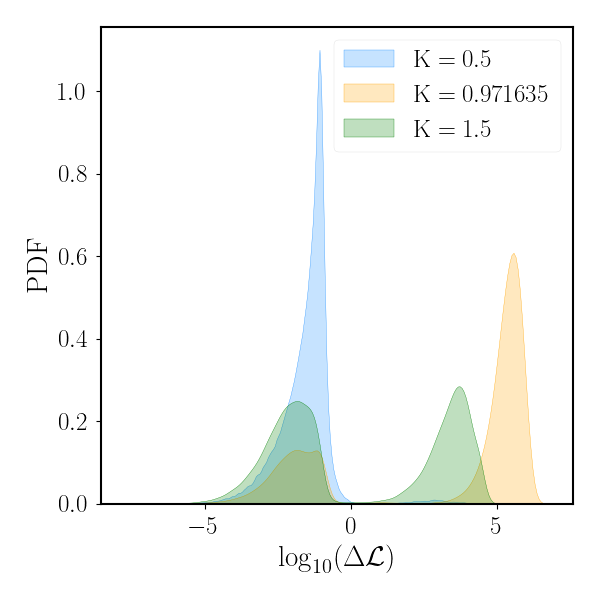}
    B) \includegraphics[scale = 0.42]{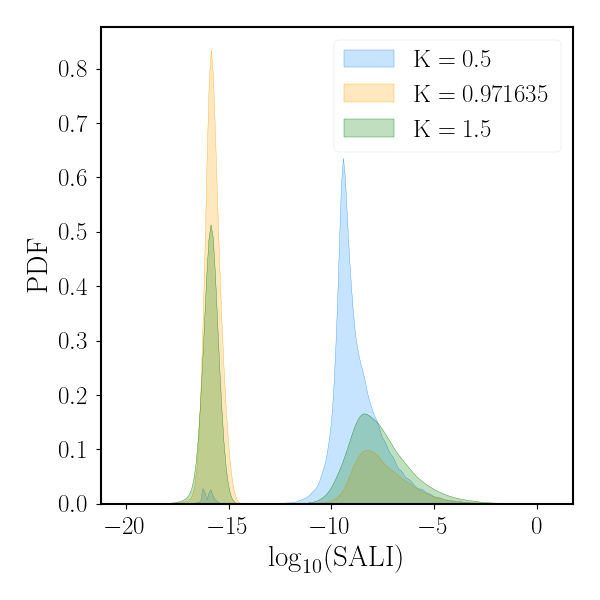} \\
    C) \includegraphics[scale = 0.42]{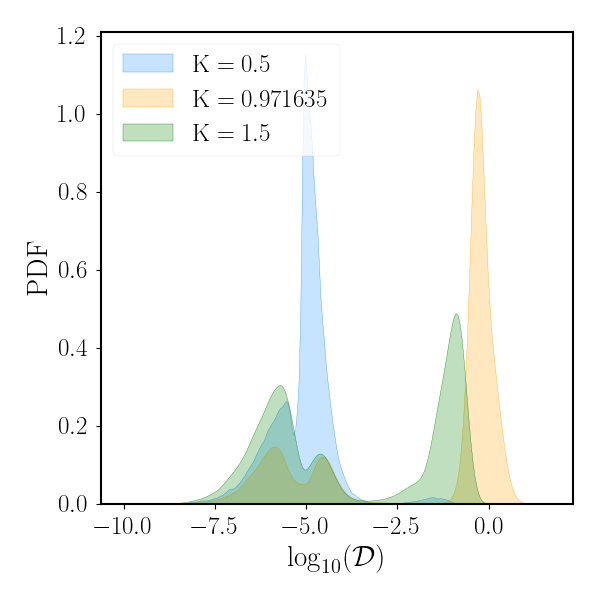}
    D) \includegraphics[scale = 0.42]{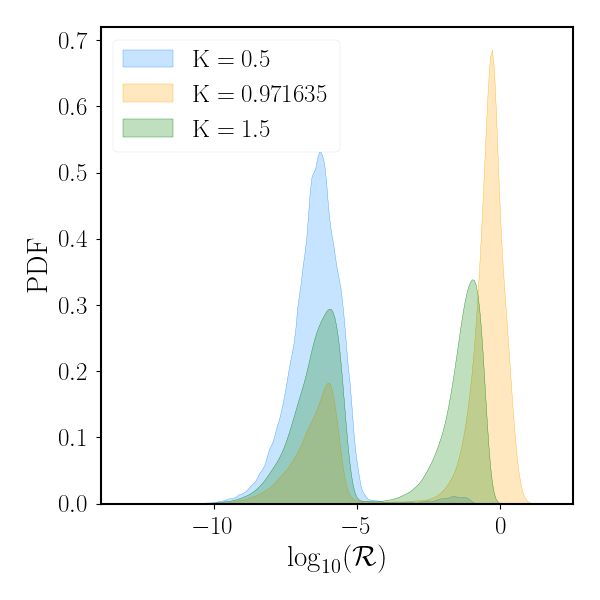} \\
    E) \includegraphics[scale = 0.42]{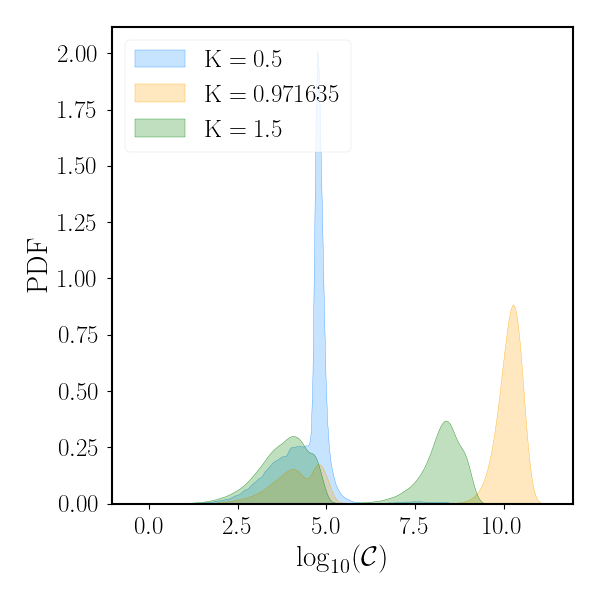}
    F) \includegraphics[scale = 0.42]{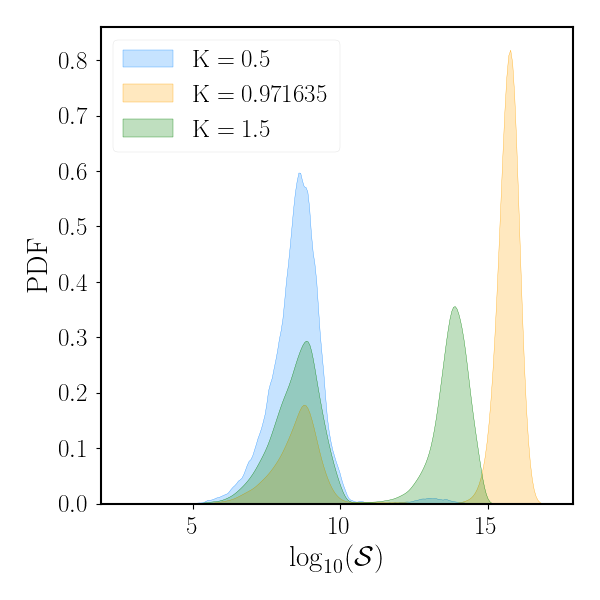}
    \caption{Histograms for the different chaos indicators considered in this analysis for the Chirikov Standard Map calculated with a regular grid of size $320 \times 320$ initial conditions and $\mathrm{K} = 0.5 , 0.971635 \,  \text{and} \, 1.5$ iterated for $N = 10^{5}$ iterations and $\| \boldsymbol{\beta} \| = \sigma = 10^{-8}$. A) distribution of the $\log_{10}(\Delta \mathcal{L})$ indicator; B) distribution of the $\log_{10}(\text{SALI})$ indicator; C) distribution of the $\log_{10}(\mathcal{D})$; D) distribution of the $\log_{10}(\mathcal{R})$; E) distribution of the $\log_{10}(\mathcal{C})$; F) distribution of the $\log_{10}(\mathcal{S})$. Note that the indicators given in Eq. \eqref{eq:chaos_inds} were calculated with $n = 2$.}
    \label{fig:histograms_2}
\end{figure}

\begin{figure}[htbp]
    \centering
    A) \includegraphics[scale = 0.42]{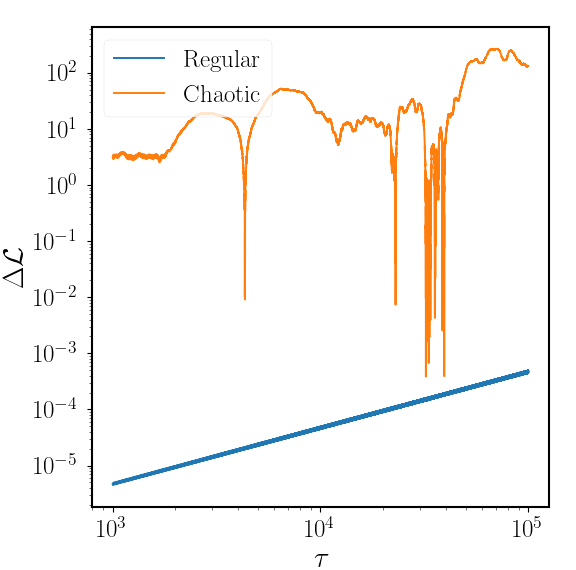}
    B) \includegraphics[scale = 0.42]{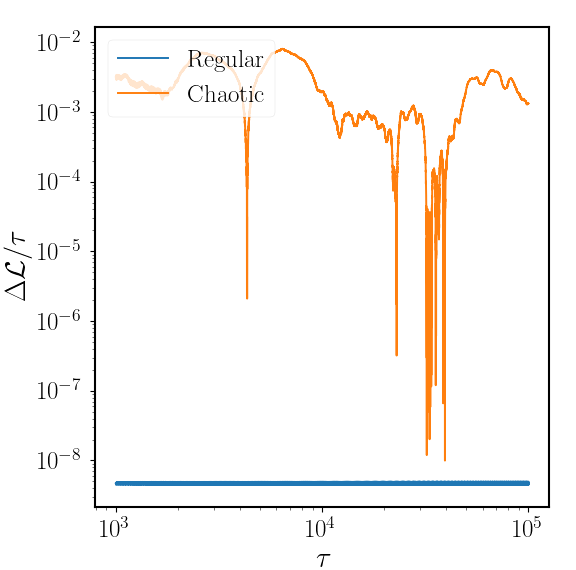}
    \caption{A) Comparison of the time evolution of the $\Delta \mathcal{L}$ indicator as a function of the integration time $\tau$ for a regular (blue) and a chaotic (orange) initial condition in the H\'enon--Heiles system with $\mathcal{H} = 1/8$. B) Comparison of the time evolution of the time-averaged $\Delta \mathcal{L}$ for the same initial conditions. As expected, for the regular initial condition the value is smaller than for the chaotic one and is approximately constant, while the evolution in the case of the chaotic trajectory is clearly different from a constant. The regular initial condition corresponds to $x = 0$, $y = 0.2$, and $p_y = 0$, while the chaotic one is given by $x = 0$, $y = -0.175$, and $p_y = 0$. To compute the neighboring trajectory, a separation of $\| \boldsymbol{\beta} \| = 10^{-8}$ was used for both cases.}

    \label{fig:comparison_DL}
\end{figure}

In Fig. \ref{fig:histograms}, we have depicted the distribution of the different chaos indicators considered for the analysis for $10^{5}$ randomly generated initial conditions in the phase space of the H\'enon-Heiles system with $\mathcal{H} = 1/8$. We can clearly see that the histograms have two peaks, which define the regular and chaotic orbits of the system. Hence, if we are interested in distinguishing order from chaos, we need to set a threshold value, that is taken in the histogram plot at the local minimum between both peaks \cite{jimenez2024}. Note that this threshold depends not only on the dynamical system under study and on its parameters, but also on the chaos indicator chosen. By visual inspection of the histogram, it is not difficult to establish this threshold, however one needs a histogram generated by large ensembles of initial conditions to carry out this task. What we have detected is that, despite being the simplest of the indicators considered in this study, the separation between the two peaks for $\Delta \mathcal{L}$ is big enough to provide a good separation of chaotic and regular dynamics. This behavior is also observed in Fig. \ref{fig:histograms_2}, where the histograms for the same chaos indicators were done for the three values of $\mathrm{K}$ considered for the Standard Map. In Fig. \ref{fig:comparison_DL} A), we show how the time evolution of $\Delta \mathcal{L}$ looks like for a regular and a chaotic initial condition for the H\'enon-Heiles system with $\mathcal{H} = 1/8$. In it, it can be seen how the growth is different for both, feature that will provide the clear separation between the peaks in the histograms for the selection of the threshold. Additionally, Fig. \ref{fig:comparison_DL} B) shows the time-evolution of the time-averaged $\Delta \mathcal{L}$ indicator. For the regular trajectory (blue), the time evolution is approximately constant, which is in good agreement with Eq. (\ref{eq:Delta_L_avg}). In the case of the chaotic trajectory, depicted in orange, the behavior is, at short times, dominated by the linear term in the polynomial form of the exponential function as higher order terms are suppressed by higher powers of $\lambda_{\text{max}}$, which for the trajectory shown is around $0.04$. Despite this, as $\tau$ grows larger, the exponential divergence will dominate and make $\Delta \mathcal{L}$ diverge.

\begin{figure}[htbp]
    \centering
    A) \includegraphics[scale = 0.5]{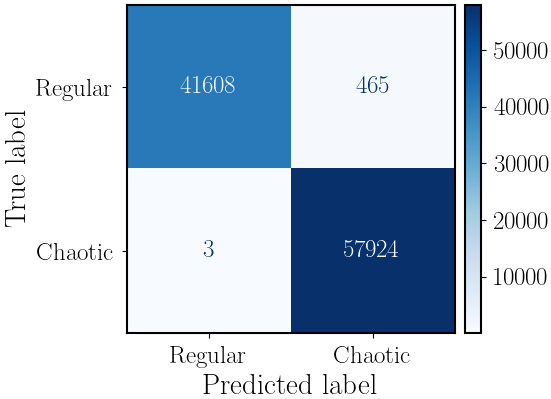}
    B) \includegraphics[scale = 0.365]{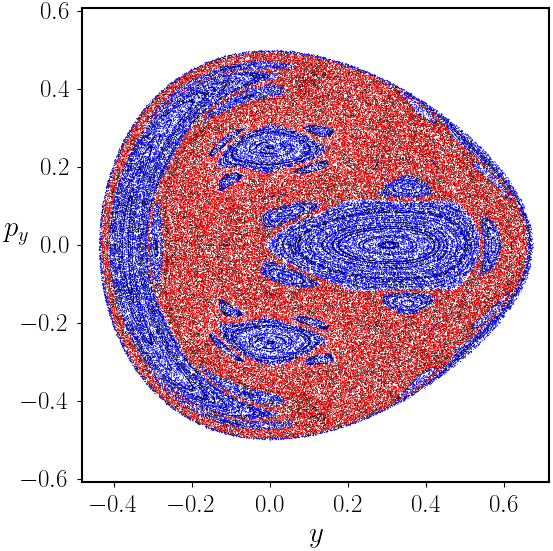} \\
    C) \includegraphics[scale = 0.5]{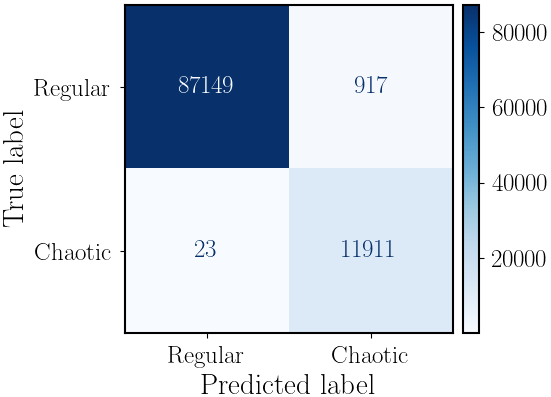}
    D) \includegraphics[scale = 0.365]{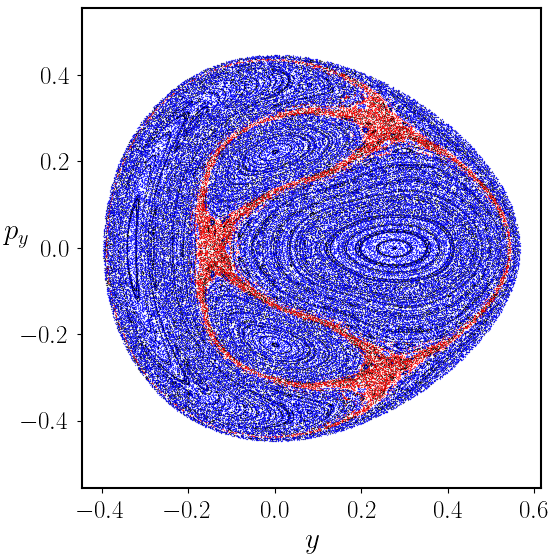} \\
    E) \includegraphics[scale = 0.5]{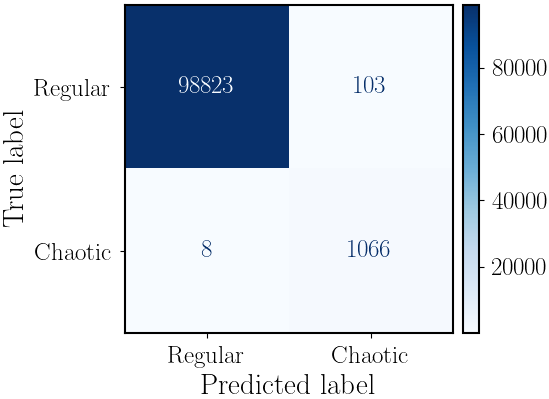}
    F) \includegraphics[scale = 0.365]{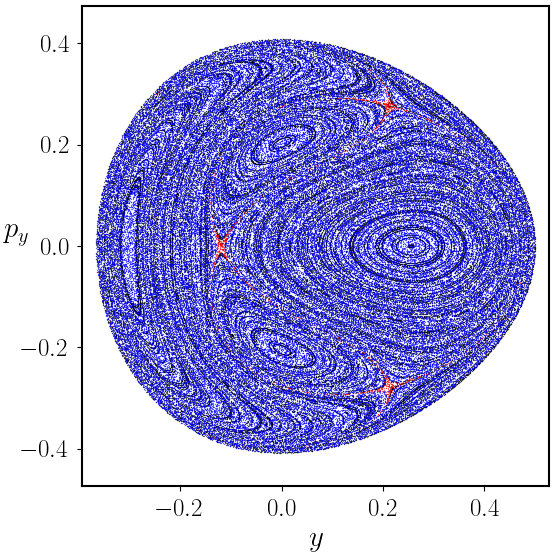} \\
    G) \includegraphics[scale = 0.5]{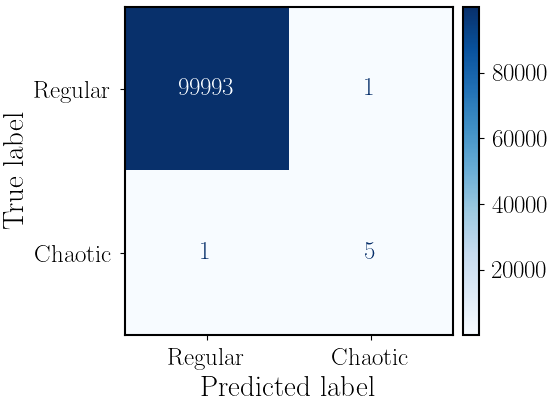}
    H) \includegraphics[scale = 0.365]{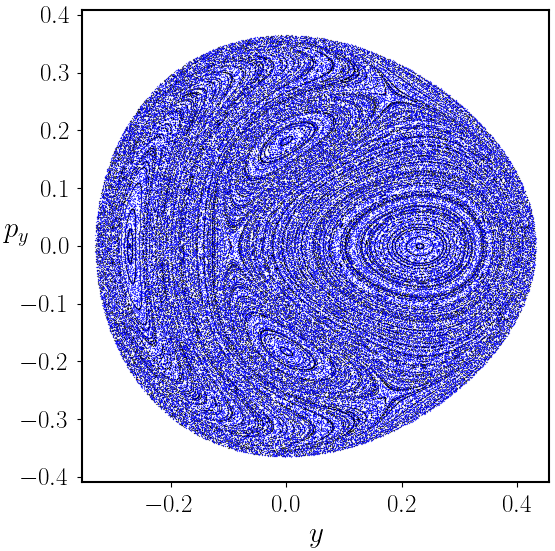}
    \caption{Comparison of orbit classification performance between $\Delta \mathcal{L}$ and SALI indicators for the H\'enon-Heiles system for various energy values. A), C), E) and G) are the confusion matrices comparing the classifications obtained with $\Delta \mathcal{L}$ and SALI for the energies $\mathcal{H} = 1/8$, $1/10$, $1/12$ and $1/15$. Here, True label stands for the classification provided by SALI while Predicted label refers to the classification obtained with $\Delta \mathcal{L}$. B), D), F) and H) are the corresponding Poincar\'e sections. For each section, $10^{5}$ randomly generated initial conditions classified with the $\Delta \mathcal{L}$ indicator, integrated for $\tau = 10^{4}$ units of time and $\|\boldsymbol{\beta}\| = 10^{-8}$, are overlaid. Regular orbits are depicted in blue and chaotic orbits in red.}
    \label{fig:HH_results}
\end{figure}

\begin{figure}[htbp]
    \centering
    A) \includegraphics[scale = 0.5]{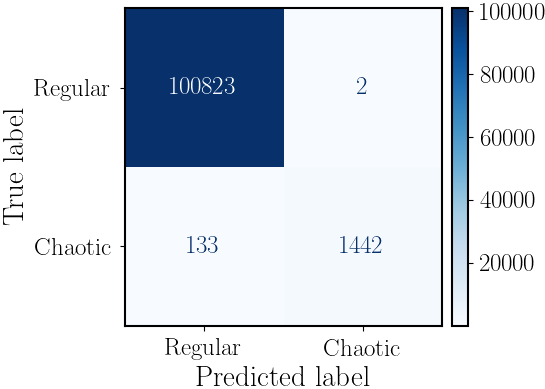}
    B) \includegraphics[scale = 0.365]{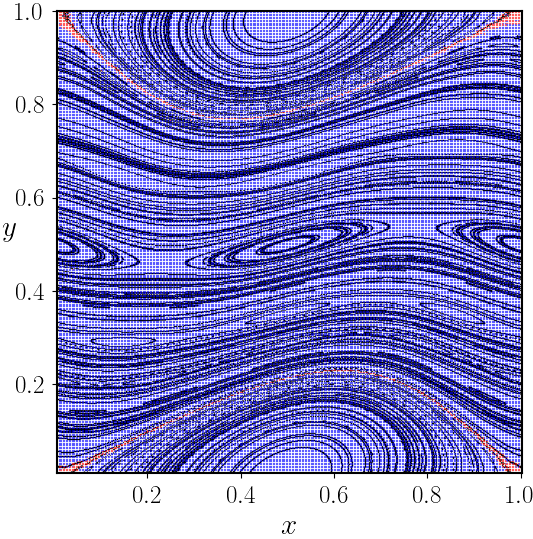} \\
    C) \includegraphics[scale = 0.5]{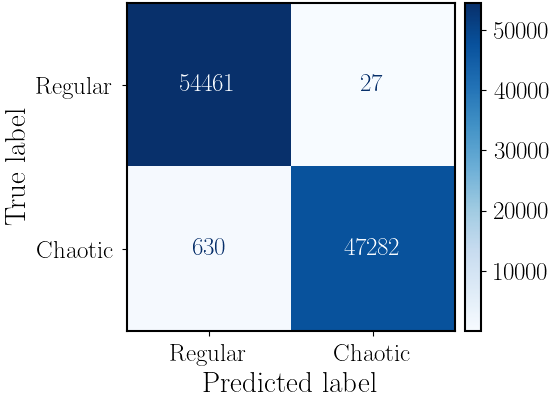}
    D) \includegraphics[scale = 0.365]{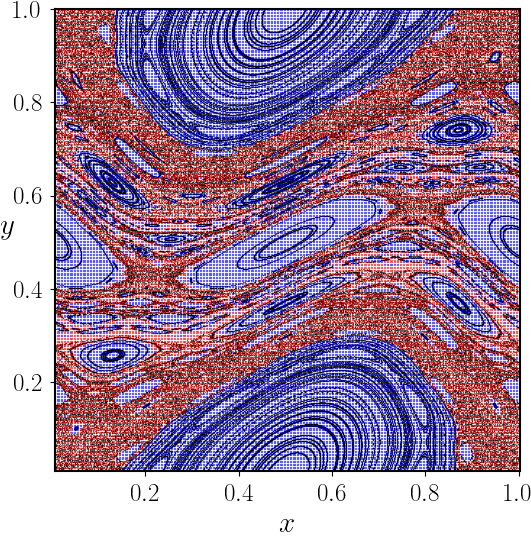} \\
    E) \includegraphics[scale = 0.5]{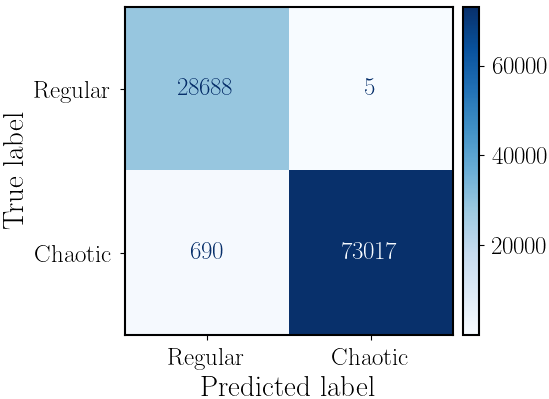}
    F) \includegraphics[scale = 0.365]{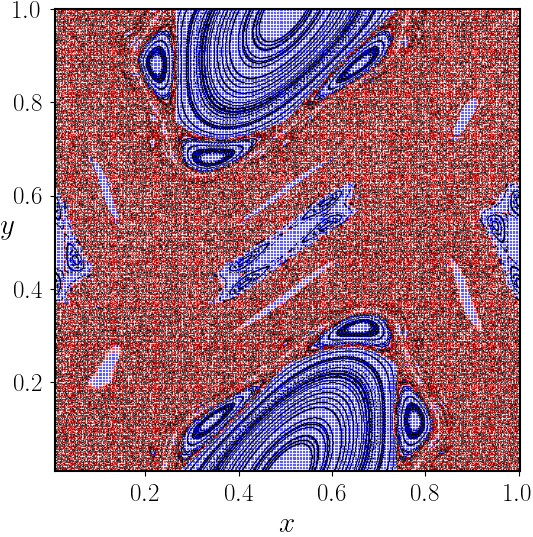}
    \caption{Comparison of orbit classification performance between $\Delta \mathcal{L}$ and SALI indicators for the Chirikov Standard Map at different $\mathrm{K}$ values. Figures A), C) and E) are the confusion matrices comparing the classification obtained with both indicators for $\mathrm{K} = 0.5$, $0.971635$ and $1.5$. Here, True label stands for the classification provided by SALI while Predicted label refers to the classification obtained with $\Delta \mathcal{L}$. B), D) and F) are the corresponding Poincar\'e sections for those $\mathrm{K}$ values. For each section, we overlay a uniform $320\times 320$ grid of initial conditions, iterated for $10^{5}$ steps and classified via the $\Delta\mathcal{L}$ indicator with $\lVert\boldsymbol{\beta}\rVert = 10^{-8}$. The ones shown in blue are classified as regular and the ones in red as chaotic.}
    \label{fig:SM_results}
\end{figure}

\begin{figure}[htbp]
    \centering
    A) \includegraphics[scale = 0.31]{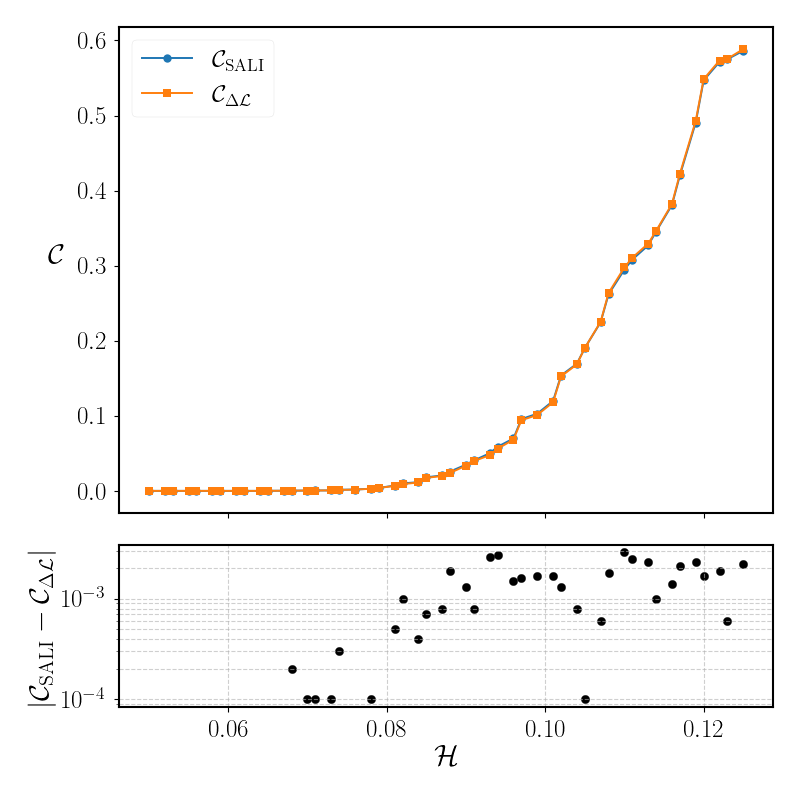}
    B) \includegraphics[scale = 0.31]{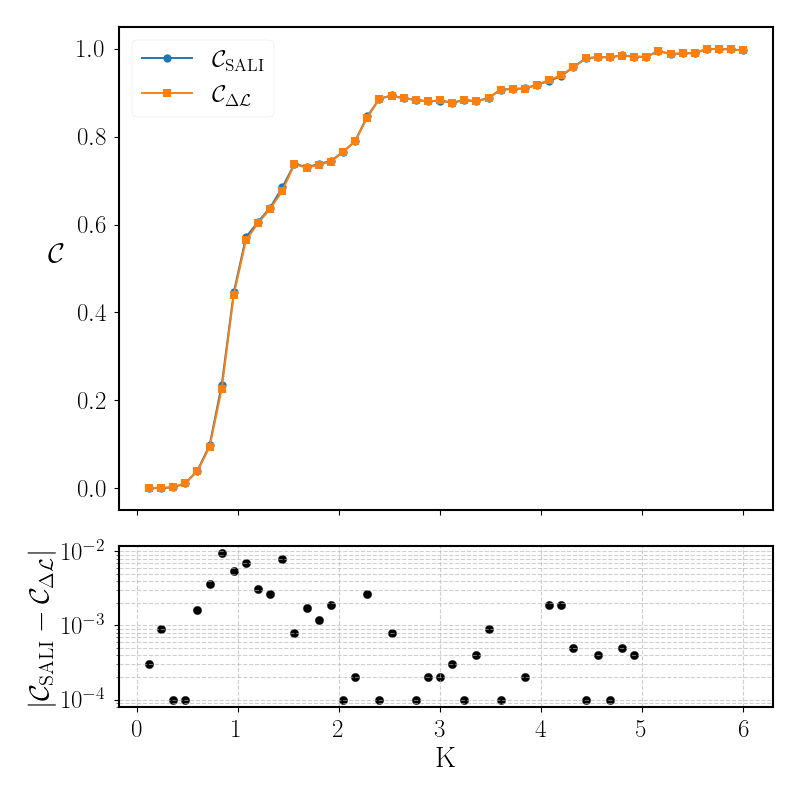}
    \caption{Comparison of the chaos fraction obtained with $\Delta \mathcal{L}$ $(\mathcal{C}_{\Delta \mathcal{L}})$ and SALI $(\mathcal{C}_{\text{SALI}})$ for: A) the H\'enon-Heiles system and B) the Chirikov Standard Map. For case A), $10^{4}$ randomly generated initial conditions, integrated for $\tau = 10^{5}$ units of time, were considered for each energy value. These values were generated as $50$ equally space values between $\mathcal{H} = 0.05$ and $\mathcal{H} = 0.125$. The threshold for SALI was set to be $\log_{10}(\text{threshold}_{\text{SALI}}) = - 8$, while for the $\Delta \mathcal{L}$ indicator it was $\log_{10}(\text{threshold}_{\Delta \mathcal{}L}) = - 1.36$ for all energy values. For B), a regular grid of $100 \times 100$ initial conditions, iterated for $N = 10^{5}$ iterations, was considered for each value of $\mathrm{K}$, whose values were obtained as $50$ equally spaced values ranging from $\mathrm{K} = 0.122$ to $\mathrm{K} = 6$. The threshold for SALI was fixed at $\log_{10}(\text{threshold}_{\text{SALI}}) = - 13$ and for the $\Delta \mathcal{L}$ indicator it was $\log_{10}(\text{threshold}_{\Delta \mathcal{}L}) = - 0.05$ for all $\mathrm{K}$ values. Note that for both systems $\|\boldsymbol{\beta}\| = 10^{-8}$.}
    \label{fig:chaos_frac}
\end{figure}

In Fig. \ref{fig:HH_results} we have calculated the confusion matrices for the test cases considered for the H\'enon-Heiles system along side the Poincar\'e sections that correspond to those energies. A confusion matrix is a table used to evaluate the performance of a classification algorithm, as it provides a detailed summary of how the prediction made by the model compares to the actual values. In our case, the rows of the matrix represent the true labels (the classification as regular or chaotic provided by SALI), while the columns correspond to the predicted labels by $\Delta \mathcal{L}$. For a binary classification problem, the matrix has four components: TP (true positives), TN (true negatives), FP (false positives) and FN (false negatives). These were previously defined at the beginning of this section. Then, a sample of $10^{5}$, randomly chosen, initial conditions were classified for each case and then plotted over the Poincar\'e section to visually present the classification provided by $\Delta \mathcal{L}$. The same process was carried out for the Chirikov Standard Map as show in Fig. \ref{fig:SM_results}. Lastly, we computed the chaos fraction as a function of the energy for H\'enon-Heiles in Fig \ref{fig:chaos_frac} A) and as a function of $\mathrm{K}$ for the Chirikov Standard Map in Fig. \ref{fig:chaos_frac} B) using $\Delta \mathcal{L}$ and SALI for $50$ values of the model parameters. In both cases, SALI and $\Delta \mathcal{L}$ produced very close results, highlighting the great performance of $\Delta \mathcal{L}$, especially if one considers the simplicity and computational advantage that this new chaos indicator introduces.

\section{Conclusions} 
\label{sec:Conclusions}

In this paper we have introduced $\Delta \mathcal{L}$, a chaos indicator derived from the method of Lagrangian descriptors that is defined as the difference in LD values between two neighboring trajectories. Its computation is remarkably straightforward, requiring only the comparison of LD values between a given initial condition and a randomly chosen infinitesimally close neighbor. We have heuristically demonstrated that $\Delta \mathcal{L}$ is upper bounded by a function that grows linearly with time for regular trajectories, while being exponentially upper bounded for chaotic ones. To assess its effectiveness in distinguishing between regular and chaotic dynamics, we benchmarked $\Delta \mathcal{L}$ against other LD-based chaos indicators, as well as the widely used Smaller Alignment Index. Using the H\'enon-Heiles system and the Chirikov standard map as test cases, our results show that $\Delta \mathcal{L}$ performs on par with these established methods. This is particularly notable given its low computational cost, as it bypasses the need for solving variational equations, a requirement typical of SALI and other established indicators.

This simplicity makes $\Delta \mathcal{L}$ a practical and efficient tool for exploring phase space structure across a wide range of dynamical systems. In future work, we aim to extend our analysis to higher-dimensional systems, where computational efficiency becomes even more critical, and to explore the properties and performance of $\Delta \mathcal{L}$ in greater detail as a function of the different parameters that are involved in its computation.

\section*{Acknowledgments} 
The authors would like to acknowledge the fruitful discussions with Prof. Jos\'e Bienvenido S\'aez Landete from Universidad de Alcal\'a, which partially motivated this paper, and for providing us with the computational resources used in the work.


\section*{CRediT authorship contribution statement}

\textbf{Javier Jim\'enez L\'opez:} Data curation, Formal analysis, Funding acquisition, Investigation, Resources, Software, Validation, Visualization, Writing - original draft, Writing - review \& editing. \textbf{V\'{i}ctor J. Garc\'{i}a-Garrido:} Conceptualization, Data curation, Formal analysis, Funding acquisition, Investigation, Project administration, Methodology, Resources, Software, Supervision, Validation, Visualization, Writing - original draft, Writing - review \& editing.

\section*{Data availability}

The data that support the findings of this study are available from the corresponding author upon reasonable request.

\bibliography{referencias}

\end{document}